\documentclass[12pt]{iopart}
\usepackage{graphicx}
\usepackage[amssymb,cdot]{SIunits}
\usepackage{xspace}

\newcommand{\STO}
	{SrTiO$_3$\xspace}
\newcommand{\AlO}
	{Al$_2$O$_3$\xspace}

\begin{document}
\title{High-$\kappa$ field-effect transistor with copper-phthalocyanine} 
\author{F. Roth and M. Huth}
\address{Physikalisches Institut, Goethe University, Max-von-Laue-Str.~1, D-60438 Frankfurt am Main, Germany}
\ead{michael.huth@physik.uni-frankfurt.de}
\begin{abstract}
The use of \STO dielectrics as high-permittivity insulator in organic thin film field effect transistors (FET) is evaluated. Field-effect transistors with sputtered \STO and copper-phthalocyanine (CuPc) as semiconducting layer were fabricated. The device preparation was performed \textit{in-situ} in an ultra high vacuum chamber system. The dielectric in the transistors had a permittivity of up to 200 which led to low driving voltages of $\unit{3}{\volt}$. The field effect transistors were \textit{p}-type and reached mobilities of about $\mu=\unit{1.5\cdot\power{10}{-3}}{\centi\square\meter\per\volt\second}$ and an on/off ratio of $10^3$. These properties are compared to devices based on other dielectric materials. 
\end{abstract}
\maketitle 
\section{Introduction}
The gate dielectric plays a crucial role for the functionality of thin film field-effect transistors. It is responsible for the reliability of the device, governs the required driving voltages and also limits the polarizations which can be achieved at the interface. In organic thin film transistors the most common dielectric is SiO$_2$. It is well known and characterized from its use in the conventional semiconductor industry, and has the advantages of excellent insulation with low leakage currents and chemical stability. The required silicon wavers with an oxide dielectric layer are commercially available and the source and drain electrodes can be easily fabricated by lithography or by deposition through a shadow mask. Various groups have in this way prepared transistors with different organic layers such as pentacene or copper-phthalocyanine \cite{Bao1996,Nelson1998}. This approach allows to concentrate on the growth of the active organic layers, but limits the experiments to the regime of low charge carrier concentrations. Many organic materials show electronic interactions which are strongly influenced by the charge carrier concentration. Examples are charge transfer salts based on organic molecules such as tetrathiofulvalene-tetracyanoquinodimethane (TTF)-(TCNQ) \cite{Torrance1979,Sarkar2010,Solovyeva2011} or K$^{+}$TCNQ$^-$ \cite{Zhou1997} which show the behavior of a Peierls or Mott insulator \cite{Mott1990}. Large changes in the charge carrier concentration may even induce a Mott metal-insulator transition, allowing for a so called Mott transition field-effect transistor (MTFET) \cite{Zhou1997,Newns1998}. It is therefore interesting to grow these materials in transistor structures where high polarizations can be reached. But also for other organic semiconductors, where no Mott metal-insulator transition occurs, the utilization of high kappa dielectrics is interesting, because the high polarizations result in transistors with low driving voltages.\\
The charge carrier concentration at the insulator surface which can be induced in field-effect experiments is 
the product of breakdown field and dielectric constant evaluated at the breakdown field \cite{Walkenhorst1992}. It therefore depends on the intrinsic properties of the material. With SiO$_2$ as dielectric a breakdown field of $E_{BD}=\unit{10}{\mega\volt\ \centi\meter^{-1}}$ and polarizations of up to \unit{3}{\micro\coulomb\ \centi\meter^{-2}} are possible at best, while with complex oxide dielectrics with a perovskite structure polarizations in the range from \unit{10-40}{\micro\coulomb\ \centi\meter^{-2}} can be achieved under carefully optimized conditions \cite{Ahn2003}.

\section{Materials}
From the perovskite class \STO is one of the best characterized materials. It has a cubic crystal structure with a lattice constant of $a=\unit{3.905}{\angstrom}$ and the advantage of a high dielectric constant of $\epsilon_{r}=300$ for bulk material at room temperature. Thin films of \STO have been grown with various techniques, such as RF sputtering \cite{Christen1994}, pulsed laser deposition \cite{Tabata1994} or molecular beam epitaxy \cite{Haeni2004}. The dielectric constant and breakdown voltage in thin film samples is in general lower than in bulk samples most likely due to defects and inhomogeneities of the electric field caused by surface roughness \cite{Christen1994}. Furthermore, \STO can be doped with niobium, which results in a conductive crystalline substrate material, ideally suited for the growth of an epitaxial dielectric \STO layer \cite{Tomio1994}.\\
In this work, to demonstrate the capability of \STO as high-$\kappa$ dielectric for organic thin film field-effect transistors, the well known organic \textit{p}-type semiconductor copper phthalocyanine was chosen as the active layer. CuPc (CuN$_8$C$_{32}$H$_{16}$) is a crystalline synthetic blue pigment from the group of phthalocyanine dyes. It has been used by various groups in field-effect experiments, usually with SiO$_2$ as the dielectric when deposited as thin films, but also already with high-permittivity insulators \cite{Bao1996,Okuda2004}.
\section{Device preparation}
An ultra high vacuum (UHV) system was employed for the preparation of the devices. It consisted of distinct chambers used for the individual process steps. The growth of the dielectric layer, active layer and the contact preparation was studied separately before they were combined to fabricate field effect transistors completely in-situ.
\subsection{SrTiO$\mathbf{_3}$ sputtering}
For the growth of the \STO layers RF-sputtering was used. The growth of sputtered \STO was characterized with X-ray diffraction employing Cu-K$_{\alpha}$ radiation in parallel beam mode and optical microscopy and optimized with respect to high crystallinity, a smooth surface and a sufficient growth rate. The sputtering target had a diameter of two inches and was at a distance of about \unit{15}{\centi\meter} from the substrate. 

\begin{figure}
\centering{
\includegraphics{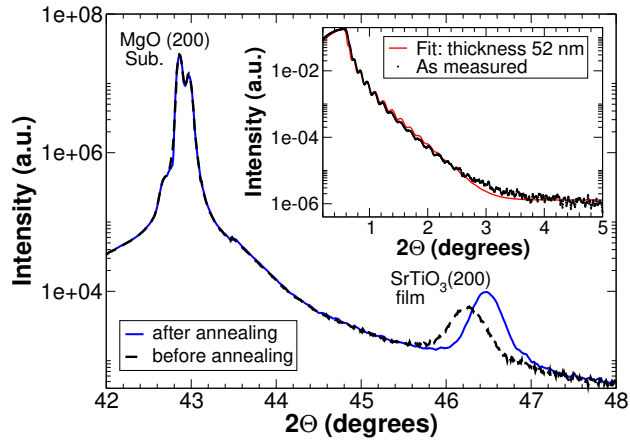}}
\caption{\label{fig:STO-combanneal}(Color online) X-ray scan with the (200) Bragg peak of a sputtered \STO film on a MgO substrate before and after annealing in air. The height of the peak increased and the center shifted toward the literature value for bulk \STO (2$\theta=46.472\,^{\circ}$). Similar behavior was observed for several films. The inset shows the corresponding thickness measurement of the \STO film and a simulated curve to determine the film thickness (\unit{52}{\nano\meter}).} 
\end{figure}

For the final FET structure preparation \STO layers had to be grown on Nb-dopend \STO (100) substrates. Since by X-ray diffraction thin film and substrate related Bragg reflexions can in this case hardly be discriminated, the dielectric layer growth was optimized on MgO (100) substrates. The thus obtained sputtering parameters were then used for FET preparation on Nb-doped \STO (100) employing an argon/oxygen mixture of 2:1 at a chamber pressure of \unit{0.025}{\milli\bbar}, a substrate temperature of \unit{720}{\degreecelsius} and a forward sputter power of \unit{100}{\watt}. The films were post annealed inside the chamber at \unit{890}{\degreecelsius} for 1 hour in an oxygen atmosphere of \unit{0.01}{\milli\bbar}. The deposition time was 90 minutes for field-effect transistor preparation which resulted in a nominal thickness of about \unit{450}{\nano\meter}. During the optimization on MgO the \STO films grew preferentially in the (100) direction. The Bragg peaks of the sputtered films were shifted to lower 2$\theta$ values as compared to bulk \STO. This was probably related to the larger lattice constant of MgO ($a=4.216$\AA) as compared to \STO ($a=3.905$\AA) which caused a tensile stress due to the misfit \cite{Peng2003}. Post annealing further improved the crystalline quality of the \STO films. This was demonstrated ex-situ in an oven at \unit{800}{\degreecelsius} for \unit{24}{\hour}, so that the sample properties could be compared before and after annealing. The \STO films showed an increase in the height of the Bragg peak and a shift in position in direction of bulk \STO, as is exemplarily shown in Fig.~\ref{fig:STO-combanneal}.
%
\subsection{CuPc growth}
\begin{figure}
\centering{
\includegraphics{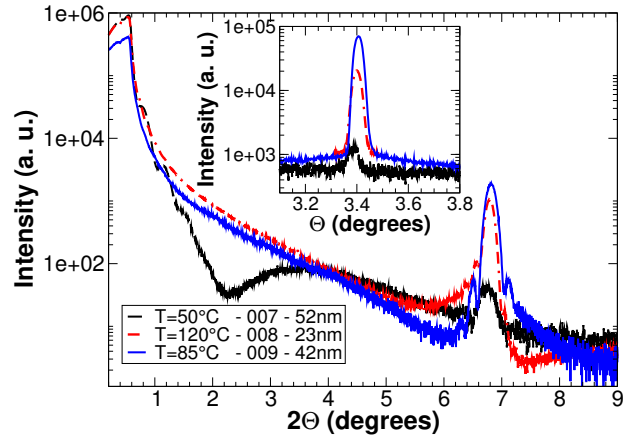}}
\caption{\label{fig:CuPc-Bragg}(Color online) Small angle X-ray diffraction pattern of the CuPc thin films prepared in the OMBD chamber on $\alpha$-\AlO substrates. The films were deposited on substrates at different temperatures. The film prepared at \unit{50}{\degreecelsius} showed pronounced Kiessig fringes but only a small Bragg peak. It was observed that this changed with higher substrate temperature - the Kiessig fringes disappeared in favor of Laue oscillations. The inset shows the corresponding rocking curves. The width (FWHM) of the rocking curve from film 009 was 0.038°. Deposition time was \unit{2}{\hour} for all films. Effusion cell temperatures were \unit{355}{\degreecelsius} for film 007, \unit{350}{\degreecelsius} for film 008 and \unit{360}{\degreecelsius} for film 009.} 
\end{figure}

\begin{figure}
\centering{
\includegraphics{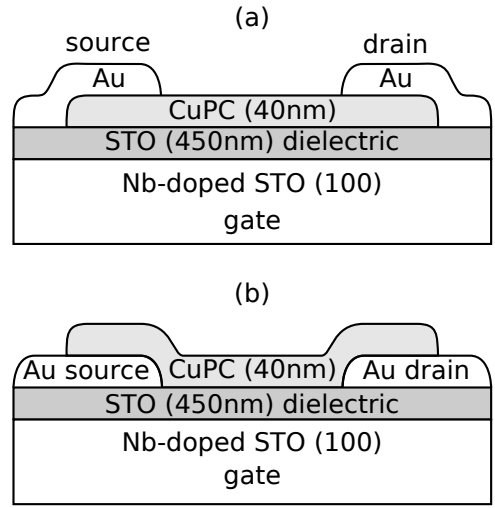}}
\caption{\label{fig:Topbottomgeom} Schematic drawing of a top contact (a) and bottom contact (b) thin film transistor with bottom gate geometry. The bottom gate is realized by a conducting substrate, here Nb doped \STO, with the dielectric layer grown on top. With the use of shadow masks both top and bottom contact devices can be realized by altering the deposition order of contact and active layer deposition.} 
\end{figure}

The organic semiconductor as active layer was prepared by organic molecular beam deposition in an UHV chamber with base pressure of \unit{\power{10}{-10}}{\milli\bbar}. The source material was commercial CuPc and it was used without further purification. The deposition rates were kept low, between \unit{0.4~and~0.02}{\nano\meter\per\min}. CuPc growth was studied at different substrate temperatures on $\alpha$-\AlO (100) and \STO (100). No difference in growth behavior between these different substrates could be observed. In contrast to this the substrate temperature had a strong impact on the growth, as observed by X-ray diffraction. The temperature influence was studied in the range from room temperature to \unit{140}{\degreecelsius}. The films changed with increasing substrate temperature from smooth, with small angle oscillations and only a small Bragg peak to higher crystallinity, with a larger Bragg peak and a rougher surface with few or no small angle oscillations. A selection of Bragg scans is shown in Fig.~\ref{fig:CuPc-Bragg} for films grown on $\alpha$-\AlO (100).
The Bragg peaks were detected for all films and substrates at an angle of 2$\theta=6.8\,^{\circ}$ which corresponds to a distance between molecular planes of \unit{1.3}{\nano\meter}. The Bragg peak is in agreement with the (200) plane of the $\alpha$-phase of CuPc. The $\alpha$-phase is frequently reported for vacuum deposition of CuPc at moderate substrate temperatures \cite{Lozzi2004}, while the $\beta$-phase can be found at high substrate temperatures or deposition rates \cite{Berger2000}. In the $\alpha$-phase the molecules are oriented close to edge-on with respect to the substrate.
For the field-effect transistors prepared in this study an intermediate substrate temperature of \unit{85}{\degreecelsius} was chosen. This temperature was selected to balance the somewhat conflicting requirements of high crystallinity and layer smoothness.\\ 
Field-effect transistors in thin film geometry with CuPc as active layer and \STO as dielectric were prepared in-situ on (100) Nb doped \STO substrates in bottom gate geometry. The source and drain electrodes were made of a \unit{30}{\nano\meter} gold layer by evaporation through shadow masks. Transistors were prepared in bottom contact (contact preparation before active layer) and top contact (contact preparation after active layer) geometry as sketched in Fig.~\ref{fig:Topbottomgeom}.
The channel length was \unit{20}{\micro\meter} and the channel width was \unit{1}{\milli\meter}. The shadow masks had a four-segment arrangement, so that there were always four (identical) transistors prepared, one in each sector of the substrate.
\section{Results}
\begin{figure}
\centering{
\includegraphics{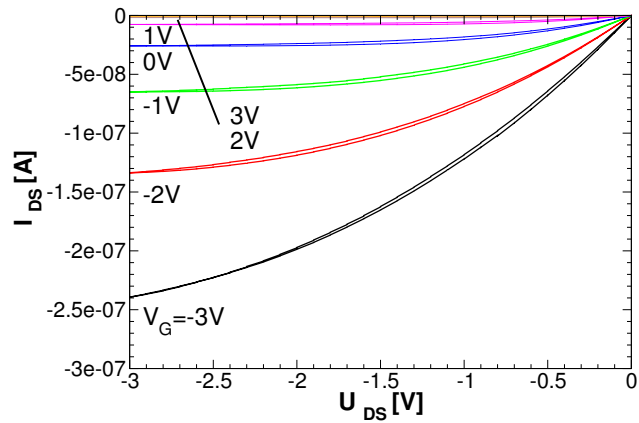}}
\caption{\label{Fig:currentvolatageSTO} Current voltage characteristic of a CuPc FET with \STO as dielectric layer. The "on" current at \unit{-3}{\volt} source drain and \unit{-3}{\volt} gate voltage was $\unit{2.4\cdot10^{-7}}{\ampere}$. The "off" current at \unit{-3}{\volt} source drain voltage and \unit{0}{\volt} gate voltage was $\unit{2.5\cdot10^{-8}}{\ampere}$ and was further suppressed to $\unit{3.5\cdot10^{-10}}{\ampere}$ with a gate voltage of \unit{3}{\volt}. The on/off ratio was about $10^3$.}
\end{figure}

\begin{figure}
\centering{
\includegraphics{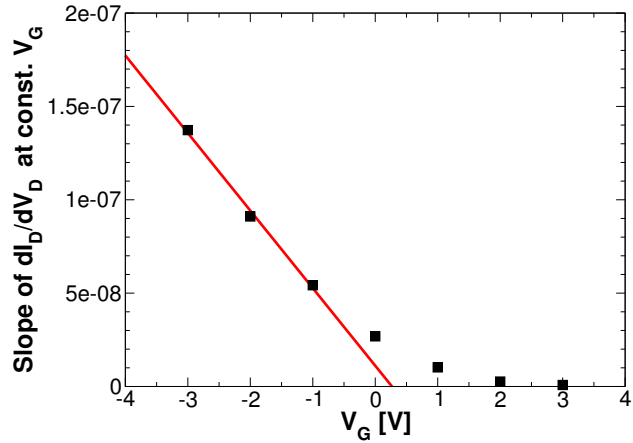}}
 \caption{\label{Fig:fet11-mobility-linear2} This plot is used to determine the mobility in the linear regime from the response curve. The slopes of the drain current over the drain voltage $dI_D/dV_D$ as a function of the gate voltage $V_G>>V_D$ (linear regime) are plotted. From the linear fit the threshold voltage was estimated by the x-axis intercept to $V_T=\unit{0.27}{\volt}$ and 
the mobility was derived from the slope to $\mu=\unit{1.5\cdot\power{10}{-3}}{\centi\square\meter\per\volt\second}$.}
\end{figure}

\begin{figure}
\centering{
\includegraphics{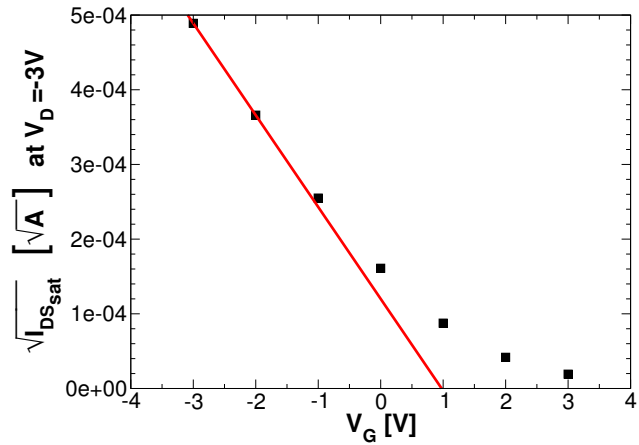}}
 \caption{\label{Fig:fet11-mobility-sat} Plot to determine the mobility in the saturation regime from
the square root of the saturation current at $V_D=\unit{-3}{\volt}$ as a function of the gate voltage. From a fit at the highest gate voltage the mobility was obtained to $\mu=\unit{1.1\cdot\power{10}{-3}}{\centi\square\meter\per\volt\second}$ and the threshold voltage to $V_T=\unit{0.98}{\volt}$.}
\end{figure}

When measured, the four FETs on one substrate showed comparable results. The dielectric layer was characterized and $I-V$ and transfer characteristics were measured ex-situ at room temperature.\\ 
The dielectric constant of the dielectric layer was calculated from the capacitance of the electrodes. The values obtained for $\kappa$ from the transistors with \STO as the dielectric layer were typically around 180 to 200 and the dielectric could usually withstand voltages of about \unit{3}{\volt} which corresponds to a breakdown voltage of about \unit{700}{\kilo\volt\ \centi\meter^{-1}} and a polarization of \unit{1.3}{\micro\coulomb\ \centi\meter^{-2}}. To calculate mobilities from the FET characteristics the capacitance per unit area was required, which was calculated from the measured capacitance and the area of the electrodes to about \unit{413}{\nano\farad\ \centi\meter^{-2}}.\\
All CuPc transistors showed \textit{p}-type semiconducting behavior. For all transistors only low voltages were necessary to achieve a pronounced modulation of the source drain current. The response characteristics of the transistors were measured multiple times with increasing voltages. The voltages were ramped up and down slowly and often a hysteresis was observed. The top contact transistors showed a better performance (i.e.~higher mobility), possibly due to an annealing effect of the active layer during contact preparation, i.~e.\ Au evaporation through a shadow mask. The response curve of a top contact FET is shown in Fig.~\ref{Fig:currentvolatageSTO}.
The transfer curves were measured for different fixed source drain voltages. The gate voltage was ramped up and down and also a hysteretic behavior was observed. A hysteresis behavior in organic FETs is not uncommon and usually related to trapped charges \cite{Mas-Torrent2004}.\\ 
For positive gate voltages the conductivity was suppressed and the drain current decreased below the zero gate voltage current, which can be interpreted as a depletion behavior. Saturation was hardly obtained for drain currents exceeding the gate voltage. This indicates a negative build in zero (threshold) voltage which biased the transistor. This FET had a on/off ratio of about $10^3$, which is comparable to literature results \cite{Okuda2004}.\\
The mobility of the transistors can be obtained in different ways. From the current voltage curve the mobility can be calculated in the saturation regime by \cite{Horowitz1998,Newman2004} 
\begin{equation}
 \mu=\left(\frac{\partial \sqrt{I_{D}}}{\partial V_{G}}\right)^2 \cdot \frac{2L}{ C_{I} W},
\end{equation}
where $L$ is the channel length, $W$ is the channel width, $C_I$ is the capacitance per unit area of the gate insulator, $I_{D}$ is the drain current and $V_{G}$ is the gate voltage. In the linear regime the mobility is obtained from the transfer curve by \cite{Horowitz1998,Newman2004} 
\begin{equation}
 \mu=\frac{\partial I_{D}}{\partial V_{G}}\cdot \frac{L}{W C_{I} V_{D}}
\end{equation}
or from the slope of a plot, where the slopes of the response curves in the linear regime at constant gate voltage are plotted over gate voltage by \cite{Horowitz1998,Newman2004}
\begin{equation} 
\mu=\frac{\partial I_{D}}{\partial V_{D}}\cdot \frac{L}{W C_{I} \left(V_{G}-V_{T}\right)},
\end{equation}
where $V_{D}$ is the drain and $V_{T}$ is the threshold voltage.
The mobility obtained in the linear regime was derived from transfer curves to $\mu=\unit{1.8\cdot\power{10}{-3}}{\centi\square\meter\per\volt\second}$. 
Additionally, the mobility was determined from the slopes of the response curves for small drain voltages. A plot of the derived slopes is presented in Fig.~\ref{Fig:fet11-mobility-linear2}.
From a fit to theses values a mobility of $\mu=\unit{1.5\cdot\power{10}{-3}}{\centi\square\meter\per\volt\second}$ and a threshold voltage of $V_T=\unit{0.27}{\volt}$ was calculated.\\
The mobility was also extracted from the saturation current as depicted in Fig.~\ref{Fig:fet11-mobility-sat}.
The currents at $V_D=\unit{-3}{\volt}$ were obtained from the response curves from Fig.~\ref{Fig:currentvolatageSTO}. The square roots of these currents were then plotted over the gate voltage and from the slope the mobility was calculated to $\mu=\unit{1.1\cdot\power{10}{-3}}{\centi\square\meter\per\volt\second}$ and the threshold voltage to $V_T=\unit{0.98}{\volt}$. The plot showed the typical increase in slope towards higher gate voltages \cite{Newman2004}.\\ 

\section{Conclusions}
In summary, FETs with \STO as dielectric and CuPc as semiconductor were prepared in-situ. The values derived for mobility in the linear and the saturation regime were in grood agreement and amounted to $\mu=\unit{1.5\cdot\power{10}{-3}}{\centi\square\meter\per\volt\second}$. These mobilities were obtained without optimization of the CuPc layer with respect to transport properties. In the literature the highest mobilities of about $\mu=\unit{1.5~to~ 2.0\cdot\power{10}{-2}}{\centi\square\meter\per\volt\second}$ were reported for a substrate temperature of \unit{125}{\degreecelsius} by Bao \textit{et al.~}\cite{Bao1996} using a SiO$_2$ dielectric and \unit{150}{\degreecelsius} by Okuda \textit{et al.~}\cite{Okuda2004} using a SiO$_2$ and a high-permittivity $\rm PbZr_{0.5}Ti_{0.5}O_3$ (PZT) insulator. Both groups report a dramatic field effect mobility increase with substrate temperature, in conjunction with higher crystalline order. At even higher substrate temperatures, the field effect mobilities decreased again, which was attributed to gaps between large non-space-filling crystals. Our field-effect mobility is well in agreement with the mobilities obtained by Bao \textit{et al.~}on SiO$_2$ at a substrate temperature of \unit{85}{\degreecelsius}. The on/off ratio ($10^3$) is comparable to the PZT devices, but lower than the best SiO$_2$ FETs ($10^5$). The driving voltages ($\unit{3}{\volt}$) are also comparable to the PZT devices ($\unit{2}{\volt}$) and much lower than the SiO$_2$ FETs ($\unit{100}{\volt}$). These results show that organic transistors with low driving voltages can be realized with a \STO dielectric. Further optimization of the \STO insulator is required to fully utilize its potential to achieve high polarizations and enable organic MTFETs, as well as basic research on the effects of large induced interface charges in organic charge transfer materials \cite{Solovyeva2009,Medjanik2010}.
\section*{Acknowledgments}
The authors are grateful for financial support through Transregio SFB TR 49 (Frankfurt, Mainz, Kaiserslautern)

\section*{References}
\begin{thebibliography}{10}
%
\bibitem{Bao1996} Z. Bao, A. J. Lovinger, and A. Dodabalapur, J. Appl. Phys. 69 (1996) 3066
%
\bibitem{Nelson1998} S. F. Nelson, Y. Lin, D. J. Gundlach, and T. N. Jackson, Appl. Phys. Lett. 72 (1998) 1854
%
\bibitem{Torrance1979} J. B. Torrance, Accounts of Chemical Research 12 (1979) 79
%
\bibitem{Sarkar2010} I. Sarkar, M. Laux, J. Demokritova, A. Ruffing, S. Mathias, J. Wei, V. Solovyeva, M. Rudloff, S. S. Naghavi, C. Felser, M. Huth, and M. Aeschlimann, Appl. Phys. Lett. 97 (2010) 111906
%
\bibitem{Solovyeva2011} V. Solovyeva and M. Huth, Synth. Met. in print (2011)
%
\bibitem{Zhou1997} C. Zhou, D. M. Newns, J. A. Misewich, and P. C. Pattnaik, Appl. Phys. Lett. 70 (1997) 598
%
\bibitem{Mott1990} N. Mott, {\em Metal-Insulator Transitions}, Taylor \& Francis, London, 1990
%
\bibitem{Newns1998} D. M. Newns, J. A. Misewich, C. C. Tsuei, A. Gupta, B. A. Scott, and A. Schrott, Appl. Phys. Lett. 73 (1998) 780
%
\bibitem{Walkenhorst1992} A.	Walkenhorst, C. Doughty, X. X. Xi, S. N. Mao, Q. Li, T. Venkatesan, and R. Ramesh, Appl. Phys. Lett. 60 (1992) 1744
%
\bibitem{Ahn2003} C. H. Ahn, J.-M. Triscone, and J. Mannhart, Nature 424 (2003) 1015
%
\bibitem{Christen1994} H.-M. Christen, J. Mannhart, E. J. Williams, and C. Gerber, Phys. Rev. B 49 (1994) 12095
%
\bibitem{Tabata1994} H. Tabata, H. Tanaka, and T. Kawai, Appl. Phys. Lett. 65 (1994) 1970
%
\bibitem{Haeni2004} J. H. Haeni, P. Irvin, W. Chang, R. Uecker, P. Reiche, Y. L. Li, S. Choudhury, W. Tian, M. E. Hawley, B. Craigo, A. K. Tagantsev, X. Q. Pan, S. K. Streiffer, L. Q. Chen, S. W. Kirchoefer, J. Levy, and D. G. Schlom, Nature 430 (2004) 758
%
\bibitem{Tomio1994} T. Tomio, H. Miki, H. Tabata, T. Kawai, and S. Kawai, J. Appl. Phys. 76 (1994) 5886
%
\bibitem{Okuda2004} T. Okuda, S. Shintoh, and N. Terada, J. Appl. Phys. 96 (2004) 3586
%
\bibitem{Peng2003} L. S.-J. Peng, X. X. Xi, B. H. Moeckly, and S. P. Alpay, Appl. Phys. Lett. 83 (2003) 4592
%
\bibitem{Lozzi2004} L. Lozzi, S. Santucci, S. L. Rosa, B. Delley, and S. Picozzi, J. Chem. Phys. 121 (2004) 1883
%
\bibitem{Berger2000} O. Berger, W. Fischer, B. Adolphi, S. Tierbach, V. Melev, and J. Schreiber, J. Mat. Sci. 11 (2000) 331
%
\bibitem{Mas-Torrent2004} M.	Mas-Torrent, M. Durkut, P. Hadley, X. Ribas, and C. Rovira, J. Am. Chem. Soc. 126 (2004) 984
%
\bibitem{Horowitz1998} G. Horowitz, J. Adv. Mat. 10 (1998) 365
%
\bibitem{Newman2004} C. R. Newman, C. D. Frisbie, D. A. da Silva Filho, J. Bredas, P. C. Ewbank, and K. R. Mann, Chem. Mater. 16 (2004) 4436
%
\bibitem{Solovyeva2009} V. Solovyeva, K. Keller, and M. Huth, Thin Solid Films 517 (2009) 6671
%
\bibitem{Medjanik2010} K. Medjanik, S. Perkert, S. Naghavi, M. Rudloff, V. Solovyeva, D. Chercka, M. Huth, S. A. Nepijko, T. Methfessel, C. Felser, M. Baumgarten, K. M\"ullen, H. J. Elmers, and G. Sch\"onhense, Phys. Rev. B 82 (2010) 245419
%
\end{thebibliography}
\end{document}